\documentclass[longauth]{aa}

\newcommand{\mstar}{$M_\star$}

\newcommand{\hb}{H$\beta$}
\newcommand{\ha}{H$\alpha$}

\newcommand{\oiii}{[O\,{\footnotesize III}]}

\newcommand{\nii}{[N\,{\footnotesize II}]}
\newcommand{\sii}{[S\,{\footnotesize II}]}
\newcommand{\kms}{km s$^{-1}$}

\newcommand{\hii}{H\,{\footnotesize II}}

\usepackage{graphicx}
\usepackage{xcolor}
\usepackage{txfonts}
\usepackage{hyperref}
\usepackage{orcidlink}

\begin{document}

\title{A merging pair of massive quiescent galaxies at $z=3.44$ in the Cosmic Vine}

   \titlerunning{A QG pair at $z=3.44$}
   \authorrunning{K. Ito et al.}

\author{K. Ito\inst{1,2,3}
\and F. Valentino\inst{1,2,4}
\and M. Farcy\inst{5}
\and G. De Lucia\inst{6,7}
\and C. D. P. Lagos\inst{8,9,1}
\and M. Hirschmann\inst{5,6}
\and G. Brammer\inst{1,10}
\and A. de Graaff\inst{11}
\and D. Bl\'anquez-Ses\'e\inst{1,2}
\and D. Ceverino\inst{12,13}
\and A. L. Faisst\inst{14}
\and F. Fontanot\inst{6}
\and S. Gillman\inst{1,2}
\and M. L. Hamadouche\inst{15}
\and K. E. Heintz\inst{1,10,16} 
\and S. Jin\inst{1,2}
\and C. K. Jespersen\inst{17} 
\and M. Kubo\inst{18}
\and M. Lee\inst{1,2}
\and G. Magdis\inst{1,2,10}
\and A. W S. Man\inst{19}
\and M. Onodera\inst{20,21}
\and F. Rizzo\inst{22}
\and R. Shimakawa\inst{23}
\and M. Tanaka\inst{20,24}
\and S. Toft\inst{1,2}
\and K. E. Whitaker\inst{15,1}
\and L. Xie\inst{25}
\and P. Zhu\inst{1,2}
}

\institute{Cosmic Dawn Center (DAWN), Denmark
\and 
DTU Space, Technical University of Denmark, Elektrovej 327, DK2800 Kgs. Lyngby, Denmark
\and
Department of Astronomy, School of Science, The University of Tokyo, 7-3-1, Hongo, Bunkyo-ku, Tokyo, 113-0033, Japan 
\and  
European Southern Observatory, Karl-Schwarzschild-Str. 2, D-85748 Garching bei Munchen, Germany
\and
\'Ecole Polytechnique F\'ed\'erale de Lausanne (EPFL), Observatoire de Sauverny, Chemin Pegasi 51, CH-1290 Versoix, Switzerland 
\and
INAF – Astronomical Observatory of Trieste, Via G. B. Tiepolo 11, 34143 Trieste, Italy
\and 
IFPU – Institute for Fundamental Physics of the Universe, Via Beirut 2, 34151 Trieste, Italy
\and 
International Centre for Radio Astronomy Research (ICRAR), M468, University of Western Australia, 35 Stirling Hwy, Crawley, WA6009, Australia.
\and
ARC Centre of Excellence for All Sky Astrophysics in 3 Dimensions (ASTRO 3D)
\and 
Niels Bohr Institute, University of Copenhagen, Jagtvej 128, 2200 Copenhagen N, Denmark
\and 
Max-Planck-Institut f\"{u}r Astronomie, K\"{o}nigstuhl 17, D-69117 Heidelberg, Germany
\and 
Departamento de Fisica Teorica, Modulo 8, Facultad de Ciencias, Universidad Autonoma de Madrid, 28049 Madrid, Spain
\and 
CIAFF, Facultad de Ciencias, Universidad Autonoma de Madrid, 28049 Madrid, Spain
\and 
Caltech/IPAC, MS 314-6, 1200 E. California Blvd. Pasadena, CA 91125, USA
\and 
Department of Astronomy, University of Massachusetts, Amherst, MA 01003, USA
\and
Department of Astronomy, University of Geneva, Chemin Pegasi 51, 1290 Versoix, Switzerland
\and 
Department of Astrophysical Sciences, Princeton University, Princeton, NJ 08544, USA
\and 
Astronomical Institute, Tohoku University 6-3, Aramaki, Aoba-ku, Sendai, Miyagi, 980-8578, Japan
\and 
Department of Physics \& Astronomy, University of British Columbia, 6224 Agricultural Road, Vancouver BC, V6T 1Z1, Canada
\and 
Department of Astronomical Science, The Graduate University for Advanced Studies, SOKENDAI, 2-21-1 Osawa, Mitaka, Tokyo, 181-8588, Japan
\and 
Subaru Telescope, National Astronomical Observatory of Japan, National Institutes of Natural Sciences (NINS), 650 North A'ohoku Place, Hilo, HI 96720, USA
\and 
Kapteyn Astronomical Institute, University of Groningen, Landleven 12, 9747 AD, Groningen, The Netherlands 
\and 
Waseda Institute for Advanced Study (WIAS), Waseda University, 1-21-1 Nishi-Waseda, Shinjuku, Tokyo 169-0051, Japana
\and 
National Astronomical Observatory of Japan, 2-21-1 Osawa, Mitaka, Tokyo, 181-8588, Japan
\and 
Tianjin Normal University, Binshuixidao 393, 300387 Tianjin, PR China}


\date{Received 28 November 2024 ; accepted 28 February 2025}

\abstract{We report the spectroscopic confirmation of a merging pair of massive quiescent galaxies at $z=3.44$. Using {\em JWST} observations, we confirm that the two galaxies lie at a projected separation of 4.5 kpc with a velocity offset of $\sim 680$ \kms\ ($\delta_z \sim 0.01$). The pair resides in the core of a known rich overdensity of galaxies, dubbed the "Cosmic Vine". For both pair members, modeling of the Spectral Energy Distributions and faint rest-frame optical emission lines indicate high stellar masses ($\log{(M_\star/M_\odot)}\sim10.9$) and suppressed star formation ($\log{\rm (sSFR/yr^{-1})}<-10$), more than an order of magnitude below the level of the star formation main sequence at this redshift. We then explore the Illustris-TNG simulation and the GAEA and SHARK semi-analytical models to examine whether they produce a pair of massive quiescent galaxies akin to that of the Cosmic Vine. While all models produce close pairs of massive quiescent galaxies at $2<z<4$ with comparable separations and velocity offsets, their predicted number densities are 10–80 times lower than our observational constraint. This discrepancy cannot be fully explained by coarse time sampling in these models or the general challenge of forming early massive quiescent galaxies in simulations. Given that $>90$\% of simulated pairs in the models that we analyzed merge by $z=0$, our findings suggest that our observed pair will likely coalesce into a single massive galaxy. The merger, occurring in the dense core of a large-scale structure, might represent a critical event in the formation of a brightest cluster galaxy and the morphological transformation of high-redshift disky quiescent galaxies into early-type ellipticals.
}

   \maketitle
\section{Introduction}
Galaxy mergers are a critical mechanism that drives the formation and evolution of galaxies. Major mergers between galaxies of similar stellar masses (mass ratio less than 1:4) can transform disky galaxies into ellipticals \citep[][]{Martin:2018:MNRAS}. In contrast, minor mergers (mass ratio greater than 1:4), especially dry minor mergers, are generally thought to drive the significant galaxy size evolution observed across time without deeply affecting their overall morphology or mass \citep[e.g.,][]{Naab:2009:ApJL, Bezanson:2009:ApJ, vanDokkum:2015:ApJ}.

In the local Universe, the majority of quiescent galaxies have bulge-like structures. On the other hand, some quiescent galaxies at high redshift, those having low star formation activity and likely progenitors of the local giant ellipticals, are found to have progressively more disk-like structures \citep[e.g.,][]{Toft:2017:Natur, Newman:2018:ApJ, Fudamoto:2022:ApJL, Ito:2024:ApJ}. This trend suggests that the morphological transformation of massive galaxies occurs after they have ceased forming stars, whose driving factor can be major mergers of quiescent galaxies. In this sense, major merging quiescent galaxies at high redshift could help us shed light on the role of merging in their morphological evolution to the local elliptical galaxies.

The occurrence of mergers likely depends on the galaxies' environment. On the one hand, in dense environments in the local Universe, such as galaxy clusters, the high-velocity dispersion makes mergers less likely to happen than in the field \citep[e.g.,][]{Merritt:1985:ApJ, Delahaye:2017:ApJ}. On the other hand, the situation might be different at high redshift. Since the progenitors of galaxy clusters (i.e., protoclusters) are often not virialized, the velocity dispersion of structures is not high, and mergers might be enhanced due to the high number of galaxies. Some observational studies indeed report higher merger fractions of massive galaxies in (proto)clusters than in the field \citep[e.g.,][]{Lotz:2013:ApJ, Hine:2016:MNRAS}. The major merger fraction as a function of the galaxy overdensity also increases in denser environments at $z\sim2-5$ \citep{Shibuya:2024:arXiv}. Moreover, recent studies report more "top-heavy" stellar mass functions in protoclusters at cosmic noon than in local clusters \citep[e.g.,][]{Sun:2024:ApJL}. This result would also imply the need for major mergers among massive galaxies to accommodate the time evolution of their number densities in clusters. Such mergers of massive galaxies in dense regions of galaxies might also contribute to forming the Brightest Cluster Galaxies observed in local cluster cores.

Recent observations have revealed the existence of physical pairs consisting of a massive quiescent galaxy and another massive galaxy, typically star-forming, at $z\sim3-5$ \citep[][]{Schreiber:2018:A&A, Kokorev:2023:ApJL, Shi:2024:ApJ}. Pairs of quiescent galaxies, in particular, are evolved systems with minimal star formation, making them ideal targets for studying the role of mergers in their post-quenching evolution. Due to the rarity of the quiescent galaxies compared to star-forming galaxies \citep[e.g.,][]{Weaver:2023:A&A}, close (e.g., $<10$ kpc in the separation) pairs of quiescent galaxies at $z>3$ have not been found. In this paper, we report the spectroscopic confirmation of a close pair of massive quiescent galaxies at $z=3.44$ using NIRSpec spectra from the {\em James Webb Space Telescope (JWST)}. This pair is located within a large-scale overdensity of galaxies, called the ``Cosmic Vine'' \citep{Jin:2024:AA}. This paper is organized as follows. In section \ref{sec:Target}, we describe the target selection. Data used in this paper are summarized in Section \ref{sec:Data}. We perform the spectral analysis, including the spectral energy distribution (SED) and the emission line fitting in Section \ref{sec:Obsresult}. In section \ref{sec:Simresult}, we compare this pair of quiescent galaxies with numerical simulations, discussing the possibility that they will merge and the possible effect of major mergers on the evolution of early massive quiescent galaxies. Our results are summarized in Section \ref{sec:Conclusion}. We assume a $\Lambda$CDM cosmology with $H_0=70\ {\rm km\ s^{-1}Mpc^{-1}}$, $\Omega_m = 0.3$, and $\Omega_\Lambda = 0.7$. Spectra and photometry used in this study are publically available \footnote{\url{https://doi.org/10.5281/zenodo.14883519}}.  

\section{Targets} 
\label{sec:Target}

\begin{figure}
    \centering
    \includegraphics[width=1\linewidth]{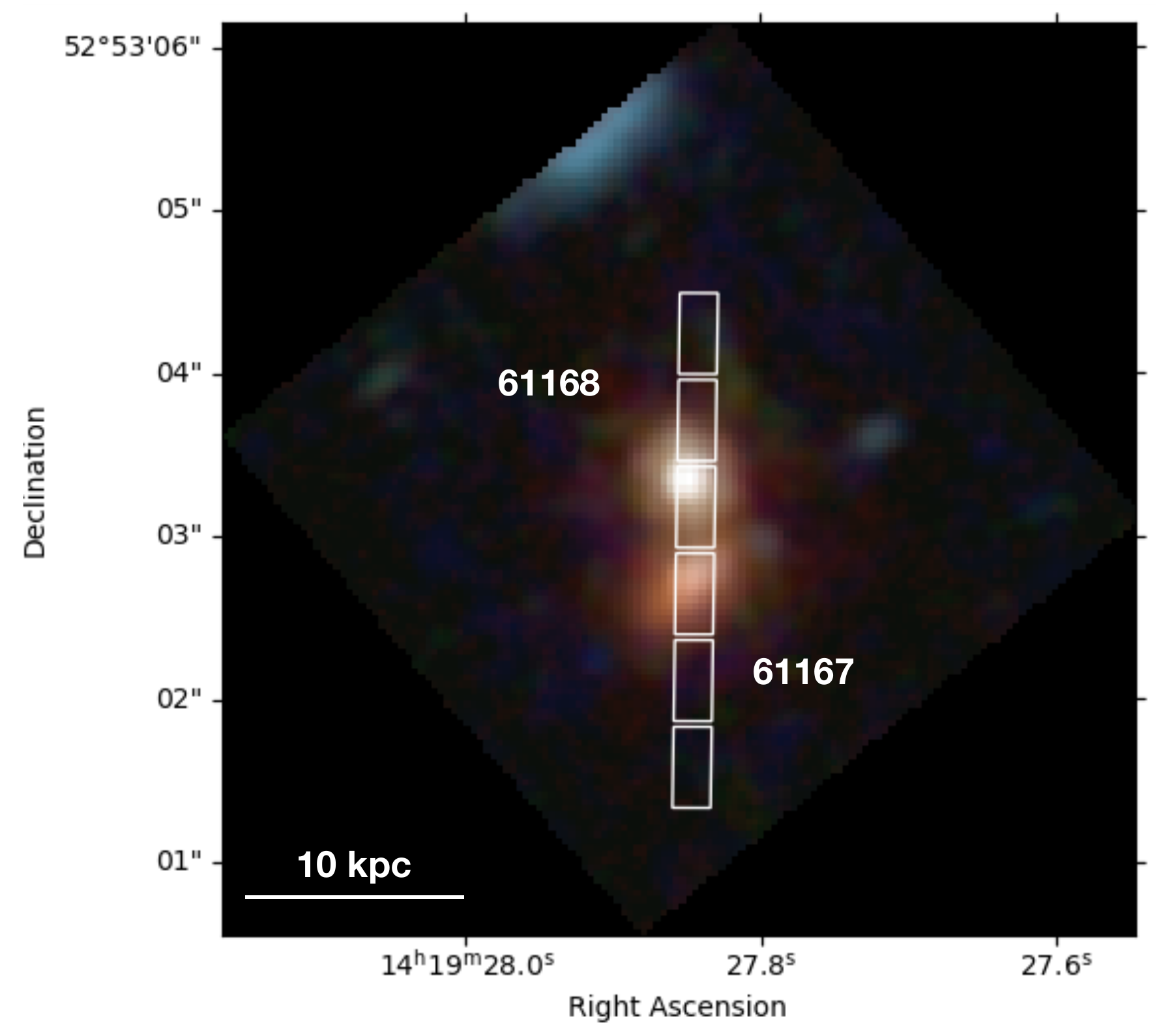}
    \caption{JWST/NIRCam color image of the pair of quiescent galaxies reported in this paper. F115W+F150W, F200W+F277W, and F356W+F444W images from the CEERS survey are used for blue, green, and red, respectively. This image is generated with  {\sc Trilogy} \protect\footnotemark. The point spread function of the blue and green images is matched to that of the F444W image. The white rectangles correspond to the open MSA shutter position.}
    \label{fig:image}
\end{figure}

Figure \ref{fig:image} shows the JWST/NIRCam image of this pair. The northern component (\#61168 in this work) of the pair has been initially $UVJ$-selected as quiescent and its spectroscopic redshift of $z_{\rm spec}=3.434$ reported as ``uncertain'' as part of a survey with Keck/MOSFIRE presented in \cite{Schreiber:2018:A&A} (ID: 3D-EGS-31322). The unambiguous confirmation has come from a {\em JWST}/NIRSpec prism spectrum in \cite{Nanayakkara:2024} (GO \#2565, PI: K. Glazebrook). This component has then fallen within the color-based selection function of the RUBIES survey \citep{deGraaff:2024:arXivb}, and targeted for prism and G395M observations with JWST/NIRSpec (GO \#4233, PI: A. de Graaff \& G. Brammer). The analysis presented here includes these observations. For what concerns the southern component (\#61167), \cite{Valentino:2023:ApJ} listed it as a candidate $UVJ$-quiescent system at $z_{\rm phot}=3.38$ and \cite{Ito:2024:ApJ} included both members in their morphological analysis\footnote{The IDs in \cite{Ito:2024:ApJ} are 9622 and 9621.}. This possible pair has been discussed in the context of its surrounding overdense environment in \cite{Jin:2024:AA}. The authors reported the existence of a massive large-scale structure elongating over $\sim4$ Mpc around the densest core hosting the pair. Approximately, there are $\sim80$ spectroscopically confirmed members to be part of this structure (Sillassen et al. in prep.). Together with the northern component, \#61167 has been targeted for G235M NIRSpec observations in a survey of quiescent galaxies at $z\sim3-4$ with the G235M grating (``DeepDive'': GO \#3567, PI: F. Valentino; Ito et al. in prep.). The spectra for both the northern and southern components are included in this work as well.

\footnotetext{\url{https://github.com/dancoe/Trilogy}}

\section{Data} \label{sec:Data}
We retrieved the available photometric data from the public DAWN {\em JWST} Archive (DJA\footnote{\url{https://dawn-cph.github.io/dja/index.html}}). NIRCam imaging was collected with the F115W, F150W, F200W, F277W, F356W, F410M, and F444W filters as part of the Cosmic Evolution Early Release Science Survey (CEERS, ERS 1345, PI: S. Finkelstein; \citealt{Bagley:2023:ApJ} for an overview). Ancillary data obtained with Hubble (with filters F606W and F814W mounted in the Advanced Camera for Surveys and the F105W, F125W, F140W, and F160W bands from the Wide Field Camera 3) were also included in the analysis. All images were homogeneously reduced with the \textsc{grizli} pipeline as described in \cite{Valentino:2023:ApJ}, including an improved treatment of the snowball effect, which is the artifact induced by a large cosmic ray impact. We adopted the aperture-corrected total photometry extracted with SEP \citep{Bertin:1996:AAS, Barbary:2016:JOSS} from mosaics in their original resolutions (thus, without prior PSF-matching) within 0\farcs5 circular apertures. We further applied the minor correction for the presence of dust in the Milky Way at this sky location \citep{Schlafly:2011:ApJ}. Noted that this photometry provides the same Spectral Energy Distribution (SED) shape as the total photometry obtained by simultaneously fitting the S\'ersic profiles to both 61168 and 61167 in \citet{Ito:2024:ApJ}.

As mentioned in Section \ref{sec:Target}, the spectra used here were acquired as part of various programs.
The data were reduced as described in \cite{Heintz:2024:arXiv} and the one-dimensional spectra optimally extracted following \cite{Horne:1986:PASP}. The spectra with the G235M grating were obtained with a single long slit covering both the 61168 and its close 61167.
Given their proximity, when combining the nodded frames for each galaxy, we masked the other pair member to avoid contamination. We conservatively increased the uncertainty by a few percent to match the sky background. We correct for remaining slit losses by anchoring the spectra to the available photometry, allowing for a smoothly varying polynomial correction (n=2) that results in flux densities increased by $\sim40$\%. No appreciable color variations are introduced at this step.\\    

\section{A pair of massive quiescent galaxies} \label{sec:Obsresult}
\subsection{SED fitting}
We simultaneously model the photometry and the available grating spectroscopy using Fast++ \citep{Schreiber:2018:A&A}. We adopt the SPS models by \cite{Conroy:2009:ApJ, Conroy:2010:ApJ} (with 3 possible stellar metallicities values of $Z=0.0096$, $0.019$, and $0.03$), a \cite{Chabrier:2003:PASP} IMF, and the \cite{Kriek:2013:ApJ} dust attenuation law (index $\delta=-0.1$, $R_{\rm V} = 3.1$, allowing for an interval of $A_{V}=0-5$ mag). Redshifts are fixed at their spectroscopic estimates, which will be derived in Section \ref{sec:specfit} and are consistent with previous photometric determinations. We use a simple delayed tau-model to parameterize the star-formation history (SFH; $\mathrm{SFR}\propto te^{-t/\tau}$ with $\mathrm{log}(\tau/{\rm yr}^{-1})=[6.5,10]$ and a minimum age of $\sim3$ Myr), which is sufficient for robust determinations of the stellar masses (\mstar) and allows for the direct comparison with ample literature. A deeper exploration of the SFHs of these systems with more complex modeling is deferred to future work (Hamadouche et al. in prep.). The best-fit parameters are reported in Table \ref{tab:tab}. We note that these results are consistent with those obtained by modeling only the photometry, and they are robust against the choice of different photometric datasets to anchor the spectra (e.g., \citealt{Weibel:2024:arXiv}). In the case of 61168, the best-fit estimates also agree with those obtained by modeling the prism spectrum in lieu of the medium-resolution grating data.

\subsection{Spectroscopic modeling}
\label{sec:specfit}
\begin{figure*}
    \centering
    \includegraphics[width=0.8\linewidth]{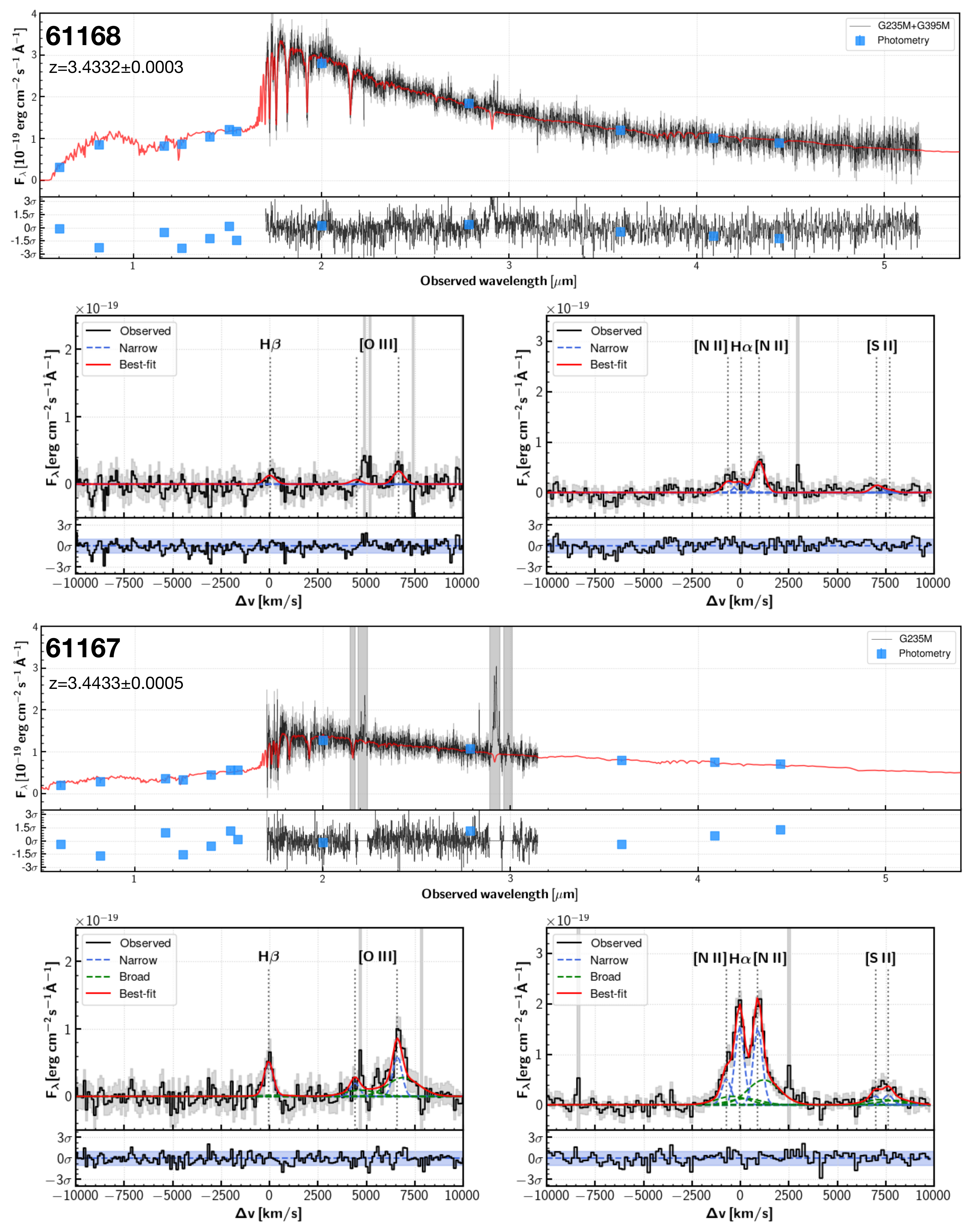}
    \caption{Summary of spectra of two quiescent galaxies in the Cosmic Vine. The first two rows represent the spectrum of the 61168, and the latter two represent the 61167. For each galaxy, the upper panel shows the observed NIRSpec medium-resolution spectrum (black line), photometry (blue squares), and the best fit in SED fitting (red line). The gray hatch indicates the wavelength masked in the SED fitting. The bottom left panel shows the continuum-subtracted spectrum around the \hb~emission line. In addition to the observed spectrum (black) and best-fit spectrum (red), the narrow and broad components of the best-fit are shown in blue and green lines, respectively. The bottom right panel shows that around the \ha~emission line. The meanings of the lines are identical to those of the bottom left panel. The difference between the observed spectrum and the best-fit model is shown at the bottom of each panel.}
    \label{fig:linefit}
\end{figure*}

We conduct line fitting, simultaneously targeting spectral features around \hb\ and \ha\ lines for these two galaxies. This is done in two steps. First, we model the stellar continuum and subtract it from the observed spectrum. Then, we conduct line fitting with the continuum subtracted spectrum. There are significant Balmer absorption lines in the stellar continuum as seen from both of their spectra at $\lambda_{\rm rest}\sim0.4\, {\rm \mu m}$ (Figure \ref{fig:linefit}). Thus, an appropriate continuum subtraction is essential to estimate line fluxes. The stellar continuum emission is modeled using {\sc ppxf} \citep{Cappellari:2017:MNRAS, Cappellari:2023:MNRAS}. We use stellar templates generated from the flexible stellar population synthesis model \citep{Conroy:2009:ApJ, Conroy:2010:ApJ}, limiting the stellar age younger than the cosmic age of the redshift of the object. A second-order multiplicative and additive correction function is used to fit the spectra. To account for the contribution of the nebular emission lines and optimally subtract the continuum, emission lines are also initially included in the model as Gaussian functions. The fitting is run in two steps. After the first pass, we mask 3$\sigma$ outliers and re-fit the spectrum, as suggested in \citet{Cappellari:2023:MNRAS}.

We then implement a more detailed modeling of the rest-frame optical emission lines in the continuum-subtracted spectra. We simultaneously model five complexes of bright lines: \hb$\lambda4861$ and \ha$\lambda6562$ in the Balmer series, and the \oiii$\lambda\lambda4959,5007$, \nii$\lambda\lambda6548,6583$, and [SII]$\lambda\lambda6716,6731$ doublets. When fitting the spectrum of the 61168, we consider only one Gaussian profile (i.e., a narrow line component) for each line. We assume no velocity offsets and the same line width among them. The line flux ratios of the \nii\ and \oiii\ doublet were fixed to 1:2.94  and 1:3, respectively \citep[e.g.,][]{Osterbrock:2006:agna}. The flux ratio of the [SII] doublet is left free to vary. This single-narrow component model poorly performed for the southern counterpart (61167), leaving significant residuals in high-velocity wings, especially in the \nii\, line (see Appendix \ref{sec:app1}). Significant residuals also remain when modeling the narrow component with different functions, e.g., a Lorentzian profile used to model the Broad emission line in active galactic nuclei (AGNs) \citep[e.g.,][]{Kollatschny:2013:A&A}. We, thus, include an additional broad component for each line, including forbidden lines. Broad components are also assumed to follow a Gaussian profile. As before, there are no velocity offsets among broad components of these lines, but we now allow for a velocity offset between narrow and broad components. Thus, we have 8 (16) free parameters in this fitting the spectrum of 61168 (61167). The fitting is conducted using the Markov Chain Monte Carlo method, and the standard deviation of the parameters is obtained from the posteriors.

Through the {\sc ppxf} fitting, we obtain the spectroscopic redshift of the southern component (61167) as $z_{\rm spec}=3.4433\pm0.0005$, which is close to that of the spectroscopic redshift of the 61168. The latter is also robustly constrained ($z_{\rm spec}=3.4332\pm0.0003$), and it is consistent with the ground-based estimate in \citet{Schreiber:2018:A&A}. We now, thus, confirm that 61168 and 61167 are physically associated given their close redshift difference ($\delta z \sim 0.01$) and small angular separation ($4.5\, {\rm kpc}$). If we convert their redshift difference into a velocity offset, it corresponds to $680 \pm 50\, {\rm km/s}$.

Figure \ref{fig:linefit} shows the best-fit emission line models for 61168 and 61167. Table \ref{tab:tab} summarizes the obtained values of the emission line fluxes. The \nii\ emission lines of 61168 are detected at $7\sigma$ significance. The signal-to-noise ratio for other lines is $\sim2\sigma$. All the narrow lines for 61167 are well detected at $\geq3\sigma$ significance, except for the \sii\ doublet ($\sim2\sigma$). In the best-fit spectra of 61167, the significant contribution of the broad line component is seen at \nii\ and \oiii\ lines. On the other hand, Balmer lines do not have a significant broad component flux. The Bayesian Information Criterion (BIC) for our fiducial model with a broad line component for \nii\ is significantly lower than that for the model without any broad line ($\Delta {\rm BIC} = 8$) and that for the model with a broad line fixed for \ha\ ($\Delta {\rm BIC} = 90$), supporting the existence of a broad \nii, rather than \ha\ line. This suggests that the broad components are likely due to outflowing ionized gas, and not due to the broad line region of an AGN. 

\begin{figure}
    \centering
    \includegraphics[width=0.9\linewidth]{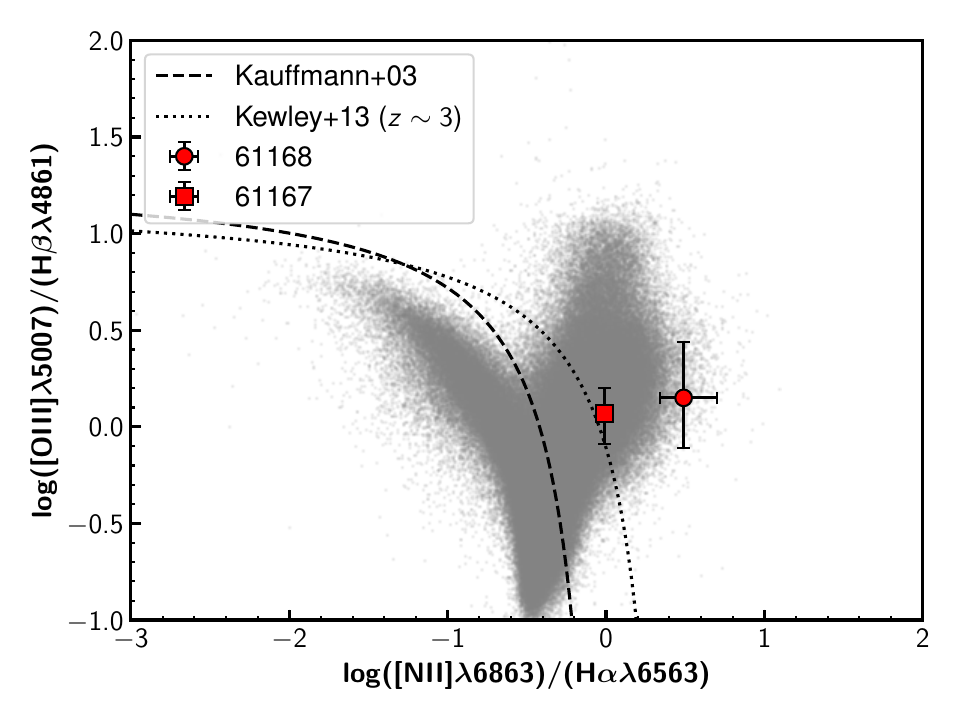}
    \caption{BPT \citep{Baldwin:1981:PASP} line ratio diagram. The pair members are shown as a red circle (61168) and a red square (61167), respectively. The lines separating the emission powered by star formation or AGN by \citet{Kauffmann:2003:MNRAS} and \citet{Kewley:2013:ApJ} are shown as dashed and dotted lines, respectively. Gray dots indicate galaxies at $z=0$ from the MPA-JHU SDSS catalog \citep{Brinchmann:2004:MNRAS,Kauffmann:2003:MNRAS,Tremonti:2004:ApJ}.}
    \label{fig:BPT}
\end{figure}

Based on classical diagnostic line ratio diagrams calibrated in the local Universe \citep[e.g.,][]{Baldwin:1981:PASP, Kauffmann:2003:MNRAS}, the \oiii${\rm \lambda5007/H\beta}$ and \nii${\rm \lambda6583/H\alpha}$ ratios (Table \ref{tab:tab}) suggest a contribution of AGNs to the observed line fluxes (see Figure \ref{fig:BPT}). The ratios also correspond to an AGN classification using the redshift-dependent parametrization of \citet{Kewley:2013:ApJ}. We note that such high line ratios have been observed in other quiescent galaxies at $z\sim2-5$ \citep[e.g.,][]{Kriek:2009:ApJL, Belli:2019:ApJ,deGraaff:2024:arXiv}.

We derive the dust extinction using the Balmer decrement from the narrow lines. Following \citet{Dominguez:2013:ApJ} and assuming a gas temperature of $T=10^4$\, K, the Case B recombination, and an electron density of $n_e=10^2\, {\rm cm^{-3}}$, we convert the \hb\ and \ha\ ratio into a color excess $E(B-V)$. The dust extinction $A_V$ is estimated from it using the reddening curve of \citet{Kriek:2013:ApJ}. Our calculations suggest the possible presence of some levels of dust extinction ($A_V\sim0-0.2$\, mag, see Table \ref{tab:tab}). We note that these values are consistent with the $A_V$ values estimated from the SED fitting, though there might be differences between the dust extinction of the nebular emission line and that of the stellar continuum light \citep[e.g.,][]{Calzetti:2000:ApJ, Kashino:2013:ApJL, Valentino:2015:ApJ}. 

Lastly, we derive the instantaneous star formation rate (SFR) based on the \ha\ fluxes of the narrow components under the extreme assumption that these photons are entirely emitted by \hii\ regions. We convert the observed \ha\ fluxes to their SFRs following \citet{Kennicutt:1998:ARA&A} adjusted for our choice of a \citet{Chabrier:2003:PASP} IMF. Considering the dust extinction with the value of $A_V$ from the Balmer decrement, the derived SFRs range within 1 to $6\, M_\odot\,  {\rm yr^{-1}}$ for both galaxies (Table \ref{tab:tab}). These values confirm that the galaxies have suppressed star formation activity ($\log{\rm (sSFR/yr^{-1})}\leq-10$), more than one dex lower than the SFR of the star formation main sequence at that redshift \citep[e.g.,][]{Schreiber:2015:A&A}. Moreover, we stress that this should be treated as an upper limit since we assume no contribution in the \ha\ emission line from the central AGN, which is unlikely according to their high-line ratios.

\begin{table}
    \caption{Summary of physical properties of two quiescent galaxies reported in this paper.}\label{tab:tab}
    \centering
    \tiny
    \begin{tabular}{ccc}
        \hline\hline
        ID & 61168 & 61167 \\
        \hline
        R.A. & 14:19:27.85 & 14:19:27.85\\
        Decl. & 52:53:03.33 & 52:53:02.72\\
        $z_{\rm spec}$ & $3.4332\pm0.0003$ & $3.4433\pm0.0005$\\
        $\log{(M_\star/M_\odot)}$ & $10.88_{-0.00}^{+0.00}$ & $10.85_{-0.00}^{+0.00}$ \\
        $\log{({\rm SFR}_{\rm SED}/M_\odot\, {\rm yr^{-1}})}$ & $<-2.0$\tablefootmark{a} & $0.48^{+0.24}_{-0.57}$ \\   
        $\log{({\rm sSFR}_{\rm SED}/{\rm yr^{-1}})}$ & $<-12.8$\tablefootmark{a} & $-10.4^{+0.2}_{-0.6}$ \\   
        $A_{V,\, {\rm SED}}$ [mag]\tablefootmark{b} & $0.34^{+0.04}_{-0.06}$ & $0.60^{+0.23}_{-0.11}$ \\
        $f_{\rm H\alpha, nr}$ [$\times 10^{-18}{\rm erg/s/cm^2}$]\tablefootmark{c} & $1.4\pm0.5 $ & $9.0\pm1.2$ \\   
        $f_{\rm H\beta, nr}$ [$\times 10^{-18}{\rm erg/s/cm^2}$]\tablefootmark{c} & $0.9\pm0.4 $ & $2.9\pm0.5$ \\   
        $f_{\rm [O\,{\footnotesize III}]\lambda5007, nr}$ [$\times 10^{-18}{\rm erg/s/cm^2}$]\tablefootmark{c} & $1.3\pm0.5 $ & $3.4\pm1.0$ \\
        $f_{\rm [N\,{\footnotesize II}]\lambda6583, nr}$ [${\times 10^{-18}\rm erg/s/cm^2}$]\tablefootmark{c} & $4.3\pm0.6 $ & $8.8\pm1.5$ \\
        $\log{\rm ([NII]\lambda6583/H\alpha)}$ & $0.49^{+0.21}_{-0.15}$ & $-0.01^{+0.05}_{-0.05}$\\
        $\log{\rm ([O\,{\footnotesize III}]\lambda5007/H\beta)}$ & $0.15^{+0.29}_{-0.26}$ & $0.07^{+0.13}_{-0.16}$\\ 
        $A_{V,\, {\rm Balmer}}$ [mag]\tablefootmark{d} & $0.00^{+0.15}_{-0.00}$ & $0.23^{+0.71}_{-0.23}$\\
        $\log{({\rm SFR}_{\rm H\alpha}/M_\odot\, {\rm yr^{-1}})}$\tablefootmark{e} & $-0.11^{+0.21}_{-0.20}$ & $0.77^{+0.32}_{-0.15}$\\
        \hline
    \end{tabular}
    \tablefoot{
    \tablefoottext{a}{The $2\sigma$ upper limit.}
    \tablefoottext{b}{This is estimated from the SED fitting.}
    \tablefoottext{c}{This is the flux of the narrow component.}
    \tablefoottext{d}{This is estimated from the Balmer decrement.}
    \tablefoottext{e}{This is obtained by assuming that the H$\alpha$ emission line is only due to the star formation and correcting the dust extinction estimated from the Balmer decrement.}
}
\end{table}

\section{Comparison with numerical simulations} \label{sec:Simresult}
The close redshift ($z=3.4332$ and $z=3.4433$) and the projected separation ($4.5\, {\rm kpc}$) suggest that this pair of massive quiescent galaxies might merge in their immediate future. Specifically, this merger is likely a major merger, given their similar stellar mass. In this section, we compare our findings with theoretical predictions. This comparison aims to understand whether simulations and semi-analytical models can produce such close pairs of quiescent galaxies and how they become at later epochs. We compare our observational results with the Illustris TNG-300 simulation \citep[$302.6^3\, {\rm cMpc^3}$ volume,][]{Nelson:2019:ComAC}, the GAlaxy Evolution and Assembly (GAEA) theoretical model \citep[$684.9^3\, {\rm cMpc^3}$,][]{DeLucia:2007:MNRAS, DeLucia:2024:A&A}, and the SHARK semi-analytical model \citep[v2, $313^3\, {\rm cMpc^3}$,][]{Lagos:2024:MNRAS}. 

\subsection{Pairs of quiescent galaxies in the numerical simulations}\label{subsec:TNG-merger}
First, we select quiescent galaxies from TNG-300, GAEA, and SHARK models. We consider the snapshots corresponding to the $2\leq z\leq4$ interval for each simulation. Galaxies are selected based on a cut in specific star formation rates ($\log{\rm (sSFR/yr^{-1})}\leq-10$). For TNG-300, this specific star formation rate is measured within twice the stellar half-mass radius, as done in our previous works \citep{Valentino:2020:ApJ, Ito:2023:ApJL}, and assuming an SFR averaged over 10\, Myr. In the case of GAEA and SHARK, the SFR is averaged over $\sim150-250\, {\rm Myr}$ and $\sim40-60\, {\rm Myr}$, respectively. Furthermore, the stellar mass cut is imposed as $\log{(M_\star/M_\odot)}\geq 10.65$ to only focus on the similar stellar mass range as the observed quiescent galaxies ($\log{(M_\star/M_\odot)}\sim 10.85$) with a margin consistent with typical systematic uncertainties (0.2 dex) on the stellar mass estimates. We note that the exact number of quiescent galaxies can differ due to the difference in the threshold of specific star formation, stellar mass, and the choice of quiescent galaxy selection \citep[e.g.,][]{Donnari:2019, Valentino:2020:ApJ}.
 
Next, we select pairs of quiescent galaxies, imposing proximity criteria on their projected sky distance ($\leq4.95$ kpc) and velocity offset ($\leq 740$ \kms). These correspond to values for the observed pair while allowing for a 10\% error margin to account for possible systematic uncertainties. We search pairs of quiescent galaxies on three projected planes (xy, yz, zx); if the quiescent galaxies are satisfied with the criteria in one of the planes, they are classified as pairs.

\begin{figure*}
    \centering
    \includegraphics[width=1\linewidth]{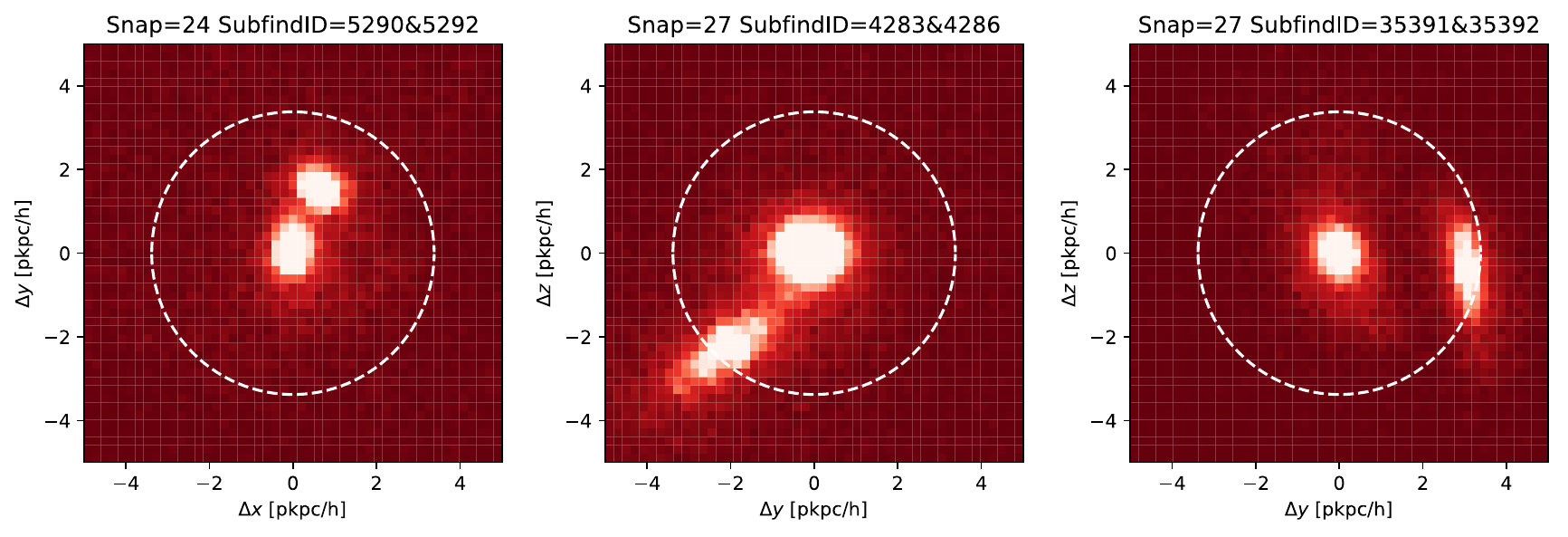}
    \caption{Stellar mass distribution of the three most distant pairs of quiescent galaxies found in the TNG-300. The left one is at $z=3.28$, and the others are at $z=2.73$. They are projected in the plane in which they are satisfied with the criteria for selecting pairs. The white circle implies the maximum separation of galaxies allowed in the criteria of pairs.}
    \label{fig:QGpair-TNG}
\end{figure*}

\begin{figure}
    \centering
    \includegraphics[width=1\linewidth]{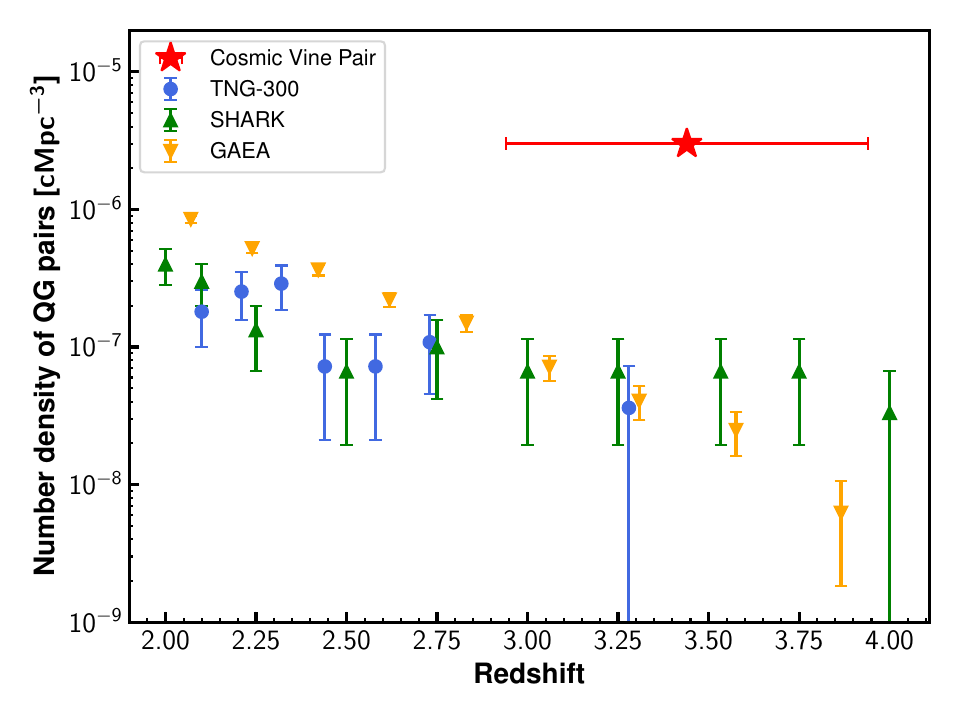}
    \caption{The number density of pairs of quiescent galaxies at each snapshot. Blue circles, green triangles, and yellow inverse triangles correspond to those of the TNG-300, SHARK, and GAEA, respectively. Their error bar is determined based on the Poisson noise. The red star shows the observationally estimated number density from the CEERS survey volume. Its error bar along the x-axis is $\pm 0.5\, {\rm Gyr}$, which is the time used in deriving the number density. }
    \label{fig:rho_QGpair}
\end{figure}

We find that all the simulations produce quiescent-quiescent pairs similar to that in the Cosmic Vine up to $z=3-4$ in terms of their separation and velocity difference. Figure \ref{fig:QGpair-TNG} shows examples of selected pairs of quiescent galaxies in TNG-300. We show the number density of these pairs as a function of the redshift in Figure \ref{fig:rho_QGpair}. Their number increases at lower redshift, which is likely largely due to the increased number of quiescent galaxies themselves \citep[e.g.,][]{Valentino:2020:ApJ, Valentino:2023:ApJ, Carnall:2023:MNRAS, Lagos:2024:MNRAS, DeLucia:2024:A&A, Nanayakkara:2024:arXiv}. The mass of host halos at $z=2$ where they are located is fairly large: more than 80\% (96\%) of pairs of quiescent galaxies in TNG-300 resided in halos with halo mass of $\log{M_{\rm halo}/M_{\odot}}>14$ ($13.5$) at $z=0$. This implies that most pairs are located in progenitors of galaxy clusters (i.e., protoclusters), as our observed pair of quiescent galaxies in the Cosmic Vine. The median pair separations are of $3.3$, $3.0$, and $1.2$\, kpc for TNG-300, GAEA, and SHARK, respectively, and their velocity offsets are $266$, $135$, and $197$ km/s. These values are smaller than our pair has ($\sim680\, {\rm km/s}$). However, there are pairs having similar offsets in each simulation, and they also have similar properties to those of the other pairs. For example, there are three pairs (corresponding 10\% of all pairs) having $>600\, {\rm km/s}$ in the TNG-300 and 29 pairs (corresponding 4\% of all pairs) in the GAEA. Here, we include all of the selected pairs in the following discussion.
 
We then turned to the available merger trees and examined whether quiescent galaxies selected as pairs will merge in the future. For TNG-300, we used the merger tree constructed with the {\sc SUBLINK} algorithm \citep{Rodriguez-Gomez:2015:MNRAS}. We find that $78\%\, (26/33)$ of pairs of QGs at $2\leq z\leq 4$ in the TNG-300 will merge within $1$ Gyr from the epoch when they are classified as pairs. $94\%\, (31/33)$ of them will merge by $z=0$. All selected pairs of quiescent galaxies in GAEA and SHARKS are also found to merge by $z\sim0$. This high fraction of mergers implies that such close pairs of massive galaxies as the one observed in the Cosmic Vine are highly likely to merge in the future.

Finally, we attempt to constrain the comoving number density of the pairs of quiescent galaxies observationally and compare it with these in simulations. It is derived by simply dividing the number of spectroscopically confirmed pairs of quiescent galaxies (1 in this case) by the survey volume of CEERS. Assuming the redshift range of $\delta z=\pm0.5$ and using the survey area of the CEERS ($\sim100\, {\rm arcmin^2}$), we estimated it to be $\sim3\times 10^{-6}\, {\rm cMpc^{-3}}$. The sum of the number density of the pairs at $z=3.44\pm0.5$ in the TNG-300, GAEA, and SHARK is $\sim 4\times10^{-8}\, {\rm cMpc^{-3}}$, $\sim1.4\times10^{-7}\, {\rm cMpc^{-3}}$ and  $\sim2.6\times10^{-7}\, {\rm cMpc^{-3}}$, respectively. Thus, the simulations show 10-80 times fewer pairs of quiescent galaxies than our observational estimate.

A possible explanation for this discrepancy is that a fraction of possible pairs might be missed because of the coarse time sampling. The time interval between snapshots in each simulation ($\sim120-160\, {\rm Myr}$) can be significantly longer than the period in which a pair is seen at such close separation. To attempt to quantify this effect, we focused on TNG300 as an example and ran simulations of four halos hosting representative pairs of quiescent galaxies merging between $z=2$ and $z=4$ using the zoom-in technique, adopting a finer time-sampling of 18 Myr. More specifically, we focused on halos with group numbers 1057, 74, 62, and 36 at $z=2$. Our goal is to check for the typical duration of when pairs are satisfied with the distance criterion used to select galaxy pairs. 
In our four re-simulations, we keep the same resolution as TNG-300 within $1.6 R_{\rm vir}$ from the halo center (and resolution progressively degrades outwards). We also use the same galaxy formation models as in the parent simulation, as extensively described by \citet{Weinberger:2017:MNRAS, Pillepich:2018:MNRAS}. We identified 10 different periods when two galaxies satisfied our criteria for close galaxy pairs among four halos. The average duration of these periods is $\sim40\, {\rm Myr}$, shorter than the interval between snapshots of the original TNG-300 simulation. This suggests that, as predicted, the coarse time sampling of the full-box TNG-300 simulation might miss a significant fraction of pairs of quiescent galaxies. However, the expected variation should amount to a factor of $(\sim120-160)/40\sim3-4$ (the ratio between the coarse sampling and the median duration of the time window when a pair satisfies our criteria), which cannot fully explain the difference between the number densities showed in Figure \ref{fig:rho_QGpair}. 

Moreover, this discrepancy is not likely only due to the lower number density of quiescent galaxies in simulations than in observations because its difference is by a factor of 3 at most at this stellar mass range ($\log{(M_\star/M_\odot)}\sim10.9$) \citep[e.g.,][]{Lagos:2024:MNRAS}. Overall, these results might suggest the possibility that the major mergers of quiescent galaxies are not reproduced well in the simulations. We note that the overdensity of the Cosmic Vine, where this pair resides, might enhance the number density of the pair of quiescent galaxies in the CEERS field to some extent. Also, the observationally constrained number density is based on one galaxy pair only. Future observations of larger samples in protoclusters and in the field will help overcome these limitations. 

\subsection{The quiescent-quiescent pairs and mergers}

In Section \ref{subsec:TNG-merger}, we showed that large simulation boxes do produce pairs of quiescent galaxies similar to that in the Cosmic Vine. These pairs will likely merge in the future ($\sim90\%$ by $z=0$). Assuming the pair in the Cosmic Vine follows their trend, these quiescent galaxies are the most distant ones with a clear feature that suggests they will experience a major merger soon. Such major mergers of quiescent galaxies might be necessary for forming the shape of the local elliptical galaxies. The S\'ersic index ($n\sim1.7$) and axis ratio ($q\sim0.4$) estimated in \citet{Ito:2024:ApJ} suggest that the southern galaxy has a disk-like morphology. This also supports the idea that this major merger might induce a morphological transformation into a final giant elliptical. Since this pair is located in a large-scale structure, the major merger of massive galaxies might also drive their evolution to the Brightest Cluster Galaxies seen in the local galaxy clusters. Moreover, considering the recent sub-millimeter observations reporting the low gas (or dust) mass fraction of quiescent galaxies at high redshift \citep{Whitaker:2021:Natur, Suzuki:2022:ApJ}, quiescent galaxies of this pair might also have deficient gas and be a dry-major merging pair. Such dry-major mergers are reported at $z<2$ \citep[e.g.,][]{Whitaker:2008:ApJL}. Below this threshold, the importance of the dry-major mergers has been observationally explored by selecting the close projected pairs of massive galaxies \citep[e.g.,][]{Bundy:2009:ApJ}, and its contribution to the stellar mass growth of quiescent galaxies and their kinematical and morphological properties are discussed \citep[e.g.,][]{Naab:2006:ApJL}. According to these observational studies, dry major mergers contribute to the stellar mass growth and the formation of slowly rotating, boxy elliptical galaxies. This pair might prove the existence of such dry mergers as early as $z\sim3$. We emphasize that this major merger is not likely the direct cause of their star formation quenching because they were already in the quiescent phase before the major merging, even though the interaction before the major merging might be somehow related to quenching.

Interestingly, there is growing observational evidence of the existence of quiescent galaxies having a massive companion at high redshift. \citet{Schreiber:2018:A&A} report a massive dust star-forming galaxy $3.1\, {\rm kpc}$ away from a quiescent galaxy, possibly on its way towards quiescence \citep{Schreiber:2021:A&A}. Both objects are spectroscopically confirmed at $z = 3.717$. Recently, \citet{Shi:2024:ApJ} reported the spectroscopic confirmation of a pair of quiescent galaxies at $z=2.24$ with a separation of $70\, {\rm kpc}$, which they argue is likely to get merged in $\sim 1\, {\rm Gyr}$. Also, \citet{Kakimoto:2024:ApJ} report a photometrically selected star-forming galaxy $12.8\, {\rm kpc}$ away from a quiescent galaxy at $z=4.53$.  All these observations hint at the essential role of major mergers in the early post-quenching evolution of massive quiescent galaxies. Further observational investigation is required to obtain a conclusive picture of the pairs of quiescent galaxies and massive galaxies. Systematic constraints on the fraction of close major merger pairs of quiescent galaxies and the investigation of their morphological properties are critical to examining the above hypothesis on the significant contribution of major mergers on the evolution of massive quiescent galaxies. A spectroscopic campaign is essential to achieve this with precise spatial separation of pairs.

\section{Summary} \label{sec:Conclusion}
In this paper, we report the spectroscopic confirmation of a close pair of massive quiescent galaxies at $z=3.44$ residing in the core of a large-scale structure, the Cosmic Vine. The pair members are close both in projected distance ($\sim4.5\, {\rm kpc}$) and along the line of sight ($\Delta v\sim 680\, {\rm km/s}$). Through the spectral modeling and comparison with simulations, we have obtained the following results:

\begin{enumerate}
\item Simultaneously modeling of the SED and spectra obtained with JWST return large stellar masses ($\log{(M_\star/M_\odot)}\sim10.9$) and strongly suppressed SFRs for both galaxies ($\log{\rm (sSFR/yr^{-1})}\leq-10$). The observed ratios of faint optical rest-frame emission lines suggest powering from AGN rather than recent star formation, further supporting the quiescence of both galaxies. 

\item We search for quiescent galaxies close pairs in the Illustris TNG-300, GAEA, and SHARK simulations to compare this quiescent pair with simulated ones. These numerical simulations produce pairs of quiescent galaxies similar to that in the Cosmic Vine at $2<z<4$ regarding their spatial and velocity separation. The vast majority ($>90\%$) of the quiescent galaxy pairs in these boxes will merge by $z=0$. This also suggests that our observed pair is also highly likely in the merging phase.

\item The number density of pairs of quiescent galaxies in the simulations is found to be $10-80$ times lower than what we derive based on this observation over the $\sim 100\, {\rm arcmin^2}$ of the CEERS field. Part of the discrepancy between the models and our limit is due to the coarse time sampling in cosmological boxes and the known difficulty in producing massive quiescent galaxies in general in simulations. However, these two effects seem unable to explain this discrepancy fully, possibly pointing to an intrinsic effect connected with major mergers at high redshift.  
\end{enumerate}

Our findings further suggest the possible importance of merging in the post-quenching evolution of the first massive quiescent galaxies. In particular, such a major merger could be critical to transform their morphology to an elliptical shape. Moreover, the fact that this pair resides in a dense large-scale structure can imply that such major mergers of quiescent galaxies form the brightest cluster galaxies in the local Universe.

\begin{acknowledgements}
    We appreciate the anonymous referee for helpful comments and suggestions that improved the manuscript. This study was supported by JSPS KAKENHI Grant Numbers JP22J00495 and JP23K13141. KI is grateful for the support of the European Southern Observatory through the Science Support Discretionary Fund for the visit. FV is grateful for the support of the Japanese Society for the Promotion of Science through the Fellowship S23108. MF and MH acknowledge funding from the Swiss National Science Foundation (SNF) via a PRIMA Grant PR00P2 193577 “From cosmic dawn to high noon: the role of black holes for young galaxies”. They also thank Volker Springel for giving the necessary information and codes to produce the TNG300 re-simulations. DC is supported by the Ministerio de Ciencia, Innovaci\'on y Universidades (MICIU/FEDER) under research grant PID2021-122603NB-C21. AWSM acknowledges the support of the Natural Sciences and Engineering Research Council of Canada (NSERC) through grant reference number RGPIN-2021-03046. The data products presented herein were retrieved from the Dawn JWST Archive (DJA). DJA is an initiative of the Cosmic Dawn Center (DAWN), which is funded by the Danish National Research Foundation under grant DNRF140.
\end{acknowledgements}

%
%
\bibliographystyle{aa}
\bibliography{bibliography}

\begin{appendix}
\onecolumn
\section{Line modeling for 61167 with single Gaussian model}\label{sec:app1}
\begin{figure*}[h!]
\centering
    \includegraphics[width=14cm]{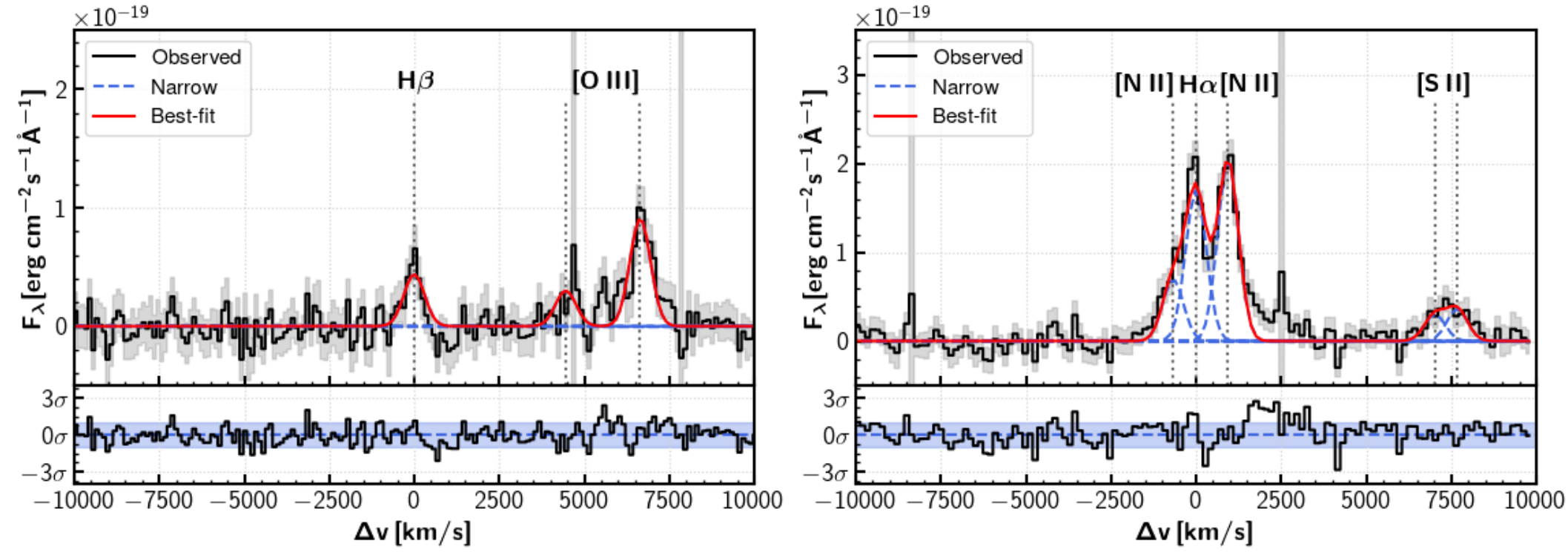}
    \caption{Summary of emission line fitting for 61167 when single Gaussian function is used for each line. The left and right panel shows the continuum-subtracted spectrum around the \hb~emission line and that around the \ha~emission line, respectively. The meanings of the lines are identical to those in Figure \ref{fig:linefit}.}
    \label{fig:app}
\end{figure*}

Figure \ref{fig:app} shows the best fit in fitting emission lines of 61167 with the single Gaussian model for each line. We see a clear residual at a redder wavelength from [N\,{\footnotesize II}]$\lambda6583$. This illustrates the necessity of additional components in the fitting.
\end{appendix}

\end{document}